\documentclass[aps,pre,preprint,groupedaddress]{revtex4-2}
\usepackage{amsfonts}
\usepackage{amsmath}
\usepackage{amssymb}
\usepackage{amsthm}
\usepackage{float}
\usepackage{graphicx}
\usepackage{bm}
\usepackage{hyperref}
\usepackage{physics}

\usepackage{lipsum} 
\usepackage{tikz}
\usepackage{stackengine}
\usepackage{wasysym}
\usepackage{xcolor}

\usetikzlibrary{decorations.pathreplacing}

\newtheorem{definition}{Definition}[section]

\usepackage[T2A]{fontenc}

\usepackage{hyphenat}
\usepackage[english]{babel}

\usepackage{color}

\begin{document}

\title{The principle of stationary action and Lagrangian for dissipative dynamics with velocity-proportional frictional force}

\author{Georgii Koniukov}
\email{kongosha@live.com}
\affiliation{Lund University, MAX IV, Lund, Sweden}

\author{Dmitry Nerukh}
\email{D.Nerukh@aston.ac.uk}
\affiliation{Department of Mathematics, Aston University, Birmingham, B4 7ET, UK}

\begin{abstract}

It has been known for a long time that the equation of motion for dissipative linear dynamical systems with constant coefficients cannot be derived from the classical principle of stationary action because the term proportional to velocity in the equation of motion leads to the time derivative of order one half in the Lagrangian.  Thus, approaches utilising fractional calculus have been used; however, they suffer from deficiencies both from mathematical and physical points of view.  We here present our version of such fractional calculus based approach that provides correct Euler-Lagrange and, ultimately, the Hamilton equations, energy change of the moving body, and an attempt for a geometric interpretation of how energy dissipates.

\end{abstract}

\maketitle

\tableofcontents
\newpage

\section{Introduction}
Since the mature formulation of the action principle by Euler, Hamilton, and Lagrange, it has been established that the equations of motion for dissipative linear dynamical systems with constant coefficients cannot be derived from a variational principle. In 1873, Lord Rayleigh introduced the Rayleigh dissipation function to handle the effects of velocity-proportional frictional forces in Lagrangian mechanics \cite{rayleigh1871}. For this, two scalar functions are required to define the equation of motion for a single system, which is not a very natural approach. Some limitations were rigorously demonstrated by Bauer in 1931 \cite{bauer1931}, who proved the impossibility of obtaining a dissipation term proportional to the first-order time derivative from such a principle. Over the years, various methods have been developed to address this issue, including the Bateman approach with auxiliary coordinates \cite{bateman1931} and time-dependent Lagrangians \cite{stevens1958}.
 Unfortunately, these methods often result in non-physical Lagrangians that fail to provide meaningful relations for the system's momentum and Hamiltonian. In 1975 Dekker \cite{dekker1975,dekker1981} introduces an innovative approach utilizing a Lagrangian that produces two first-order equations as complex conjugates, which is logical given the frequent appearance of complex numbers in the context of dissipation. These equations can be combined to yield a real, second-order equation of motion. However, it is important to note that this method is specifically applicable to a damped oscillator.

In 1996-1997, Riewe discovered a gap in Bauer's proof, which originally assumed that all derivatives were of integer order. Riewe incorporated Riemann-Liouville fractional time derivatives into the Lagrangian \cite{riewe1996}, successfully deriving equations of motion that included non-conservative forces such as friction. While innovative due to the redefinition of the action principle, this work relied on non-physical approximations and assumptions, such as inappropriately changing the limits of integration in fractional derivatives or applying an unphysical limit where the observation time interval is artificially set to zero.

In 2014, Lazo and Krumreich refined Riewe's approach by replacing the Riemann-Liouville derivative with Caputo's fractional derivative and utilizing the semigroup property of fractional integral operators \cite{lazo2014}, which made the calculations more straightforward. However, the unphysical time limit remained.
\par Despite these advances, a unified and physically meaningful method to incorporate non- conservative forces within the framework of Lagrangian mechanics is still lacking. In 2013, Galley \cite{galley2013} introduced a new formulation for Lagrangian and Hamiltonian dynamics in nonconservative systems by modifying Hamilton’s principle, thus expanding the scope of variational methods to include dissipative effects. However, this approach requires further modifications to the classical action principle, which can complicate the resulting Lagrangian and Hamiltonian structures.

Here, we propose an alternative approach that derives friction forces by still using fractional calculus. Our method avoids non-physical time limits. It provides the correct equation of motion as well as energy change.

\section{Obtaining the friction term $\gamma\dot{x}$ from a Lagrangian}
\subsection{The problem statement}
Consider a one-dimensional particle of mass $m$, with position $x(t)$, moving in a potential $U(x)$ and subject to a linear viscous friction force
\begin{equation}
    F_{\mathrm{fr}}=-\gamma\dot{x},
\end{equation}
where $\gamma>0$ is the linear friction (viscous damping) coefficient. Its equation of motion is, therefore,
\begin{equation}
    m\ddot{x}+\gamma\dot{x}
    =
    -\frac{dU}{dx}.
\end{equation}
In the absence of friction, the dynamics follows from the usual conservative Lagrangian
\begin{equation}
    L_0(x,\dot{x})
    =
    \frac{m}{2}\dot{x}^{\,2}-U(x).
\end{equation}
The question is whether the friction term $\gamma\dot{x}$ can be generated by adding a single quadratic term to $L_0$. Consider, for this purpose, a term depending on an $n$th-order time derivative,
\begin{equation}
    L_n
    =
    \frac{\lambda_n}{2}
    \left(
        \frac{d^n x}{dt^n}
    \right)^2,
\end{equation}
where $\lambda_n$ is a constant coefficient. For an integer $n$, the generalized Euler-Lagrange equation contains the contribution
\begin{equation}
    (-1)^n
    \frac{d^n}{dt^n}
    \left(
        \frac{\partial L_n}
             {\partial (d^n x/dt^n)}
    \right)
    =
    (-1)^n\lambda_n
    \frac{d^{2n}x}{dt^{2n}}.
\end{equation}
Thus, a quadratic dependence on an $n$th-order derivative in the
Lagrangian produces a derivative of order $2n$ in the equation of
motion. To reproduce linear friction, whose derivative order is one,
one would have to require
\begin{equation}
    \frac{d^{2n}x}{dt^{2n}}
    \propto
    \frac{dx}{dt},
\end{equation}
and hence
\begin{equation}
    n=\frac{1}{2}.
\end{equation}
This motivates the introduction of a
half-order fractional derivative: formally, a quadratic term of the form
\begin{equation}
    \frac{\gamma}{2}
    \left(
        D_t^{1/2}x
    \right)^2
\end{equation}
is the natural candidate for producing a first-order term
$\gamma\dot{x}$ in the equation of motion. In the next subsection, we will define such fractional derivative.
\subsection{Fractional derivatives}

The Riemann–Liouville fractional integral follows naturally from
Cauchy's formula for repeated integration. For an integer $n\geq 1$,
an $n$-fold integral can be written as
\begin{equation}
    \int_a^t x(\tilde t)\,(d\tilde t)^n
    =
    \int_a^t dt_n
    \int_a^{t_n} dt_{n-1}
    \cdots
    \int_a^{t_2} x(t_1)\,dt_1 .
\end{equation}
Changing the order of integration gives Cauchy's formula
\begin{equation}
    \int_a^t x(\tilde t)\,(d\tilde t)^n
    =
    \frac{1}{(n-1)!}
    \int_a^t
    (t-u)^{n-1}x(u)\,du .
\end{equation}
Since
\begin{equation}
    (n-1)! = \Gamma(n),
\end{equation}
this can equivalently be written as
\begin{equation}
    \int_a^t x(\tilde t)\,(d\tilde t)^n
    =
    \frac{1}{\Gamma(n)}
    \int_a^t
    \frac{x(u)}{(t-u)^{1-n}}\,du .
\end{equation}
The right-hand side is meaningful for non-integer positive values of
$n$ through the Gamma function. Replacing the integer $n$ by a real
order $\alpha>0$, therefore, leads to the left Riemann-Liouville
fractional integral
\begin{equation}
    {}_aJ_t^\alpha x(t)
    =
    \frac{1}{\Gamma(\alpha)}
    \int_a^t
    \frac{x(u)}{(t-u)^{1-\alpha}}\,du .
\end{equation}
The definitions below \cite{samko1993} continuously extend the familiar integer-order
derivatives. More generally, for a Riemann-Liouville derivative of fractional order $p$, choose an integer
$k$ such that $k-1\leq p<k$. Then
\begin{equation}
    {}_aD_t^p f(t)
    =
    \frac{d^k}{dt^k}
    {}_aJ_t^{\,k-p}f(t),
\end{equation}
where
\begin{equation}
    {}_aJ_t^{\,k-p}f(t)
    =
    \frac{1}{\Gamma(k-p)}
    \int_a^t
    (t-\tau)^{k-p-1}f(\tau)\,d\tau
\end{equation}
is the left Riemann-Liouville fractional integral of order $k-p$.

At the integer endpoint $p=k-1$, one has $k-p=1$, so that
\begin{equation}
    {}_aJ_t^{\,k-p}f(t)
    =
    {}_aJ_t^1 f(t)
    =
    \int_a^t f(\tau)\,d\tau,
\end{equation}
Then, at the integer endpoint $p=k-1$, the fractional integral becomes an ordinary integral of order one,
\begin{equation}
    {}_aD_t^{k-1}f(t)
    =
    \frac{d^k}{dt^k}
    {}_aJ_t^1f(t)
    =
    \frac{d^k}{dt^k}
    \int_a^t f(\tau)\,d\tau
    =
    f^{(k-1)}(t).
\end{equation}

\subsection{Our case: fractional derivatives of order $\frac{1}{2}$}

Let $x(t)$ be defined on a finite time interval $[a,b]$. For the fractional
order $\alpha=1/2$, the left and right Riemann-Liouville derivatives are
defined as
\begin{equation}
    {}_{a}D_{t}^{1/2}x(t)
    =
    \frac{1}{\sqrt{\pi}}
    \frac{d}{dt}
    \int_{a}^{t}
    \frac{x(\tau)}{\sqrt{t-\tau}}\,d\tau 
\end{equation}
and
\begin{equation}
    {}_{t}D_{b}^{1/2}x(t)
    =
    -\frac{1}{\sqrt{\pi}}
    \frac{d}{dt}
    \int_{t}^{b}
    \frac{x(\tau)}{\sqrt{\tau-t}}\,d\tau .
\end{equation}
Here $a$ and $b$ are the left and right endpoints of the time interval,
respectively. The derivatives are nonlocal: the left derivative depends
on the history of $x$ over $[a,t]$, whereas the right derivative depends
on its values over $[t,b]$.

For the Caputo definition, the ordinary derivative is taken before the
fractional integration. The corresponding half-order derivatives are
\begin{equation}
    {}_{a}^{C}D_{t}^{1/2}x(t)
    =
    \frac{1}{\sqrt{\pi}}
    \int_{a}^{t}
    \frac{\dot{x}(\tau)}{\sqrt{t-\tau}}\,d\tau 
\end{equation}
and
\begin{equation}
    {}_{t}^{C}D_{b}^{1/2}x(t)
    =
    -\frac{1}{\sqrt{\pi}}
    \int_{t}^{b}
    \frac{\dot{x}(\tau)}{\sqrt{\tau-t}}\,d\tau .
\end{equation}
Thus, the essential difference between the Riemann-Liouville and Caputo
definitions is the order in which ordinary differentiation and
fractional integration are performed \cite{lazo2014}.

\subsection{Problems with existing approaches}
Riewe \cite{riewe1996}  proposed, in terms of Riemann--Liouville derivatives, the Lagrangian
\begin{equation}
    L_{\mathrm R}
    =
    \frac{m}{2}\dot{x}^2-U(x)
    +i\frac{\gamma}{2}
    \left({}_{t}D_{b}^{1/2}x\right)^2 .
\end{equation}
The corresponding fractional Euler-Lagrange equation contains both
left and right fractional derivatives:
\begin{equation}
    m\ddot{x}
    -i\gamma\,
    {}_{a}D_{t}^{1/2}
    {}_{t}D_{b}^{1/2}x
    =
    -\frac{dU}{dx}.
\end{equation}
The difficulty is that, because fractional derivatives are nonlocal,
the mixed composition
\begin{equation}
    {}_{a}D_{t}^{1/2}
    {}_{t}D_{b}^{1/2}
\end{equation}
cannot in general be identified with an ordinary first derivative.
Riewe therefore supplemented the variational principle by the
short-time limit $a\to b$ and used the approximation
\begin{equation}
    i\,{}_{a}D_{t}^{1/2}f(t)
    \simeq
    {}_{t}D_{b}^{1/2}f(t),
\end{equation}
so that the two half-order operators could effectively be composed,
leading formally to
\begin{equation}
    m\ddot{x}+\gamma\dot{x}
    =
    -\frac{dU}{dx}.
\end{equation}
The leads to substantial mathematical problems: such left-right replacement does not seem to be valid, and the relevant
Riemann-Liouville expressions may become undefined when the interval
$[a,b]$ is collapsed.
\par Lazo and Krumreich \cite{lazo2014} reformulated this construction using
Caputo derivatives and obtained the Lagrangian
\begin{equation}
    L
    =
    \frac{m}{2}\dot{x}^2-U(x)
    +\frac{\gamma}{2}
    \left({}_{t}^{C}D_{b}^{1/2}x\right)^2 .
    \label{eq:fractional_friction_lagrangian}
\end{equation}
It is important to stress that the use of the Caputo derivative in the
Lagrangian still produces the fractional Euler-Lagrange equation associated with the mixed left-right operator
\begin{equation}
    {}_{a}D_{t}^{1/2}\,
    {}_{t}^{C}D_{b}^{1/2}x 
\end{equation}
and explicitly takes the form
\begin{equation}
    m\ddot{x}
    -
    \gamma\,
    {}_{a}D_{t}^{1/2}
    {}_{t}^{C}D_{b}^{1/2}x
    =
    F(x).
\end{equation}
It is still needed to take the limit $a\to b$, with
$t=(a+b)/2$, leading to
\begin{equation}
    {}_{t}^{C}D_{b}^{1/2}x
    \simeq
    -{}_{a}^{C}D_{t}^{1/2}x.
\end{equation}
The mixed operator can, therefore, be replaced by
\begin{equation}
    -{}_{a}D_{t}^{1/2}
    {}_{a}^{C}D_{t}^{1/2}x,
\end{equation}
for which both fractional operators are now left-sided.
An additional technical advantage of their formulation is that the
composition of a Riemann-Liouville derivative with a Caputo derivative
of the same orientation can be reduced exactly by using the semigroup
property of Riemann-Liouville fractional integrals,
\begin{equation}
    {}_{a}J_{t}^{\alpha}\,{}_{a}J_{t}^{\beta}
    =
    {}_{a}J_{t}^{\alpha+\beta}.
\end{equation}
In particular, for $\alpha=\beta=1/2$,
\begin{align}
    {}_{a}D_{t}^{1/2}\,
    {}_{a}^{C}D_{t}^{1/2}x(t)
    &=
    \frac{d}{dt}\,
    {}_{a}J_{t}^{1/2}
    {}_{a}J_{t}^{1/2}
    \frac{dx}{dt}
    \\
    &=
    \frac{d}{dt}\,
    {}_{a}J_{t}^{1}
    \frac{dx}{dt}
    =
    \frac{dx}{dt}.
\end{align}
Thus, once the left- and right-sided fractional derivatives appearing
in the variational equation have been brought to the same orientation,
the emergence of the ordinary first derivative is no longer merely a
formal ``$1/2+1/2=1$'' argument, but follows from the semigroup property
of the underlying Riemann-Liouville fractional integrals. \par This semigroup property hinted that the limit $a\to b$ might not be necessary. \\

Despite this progress, deriving the expression
\begin{equation}
    \frac{dE}{dt}=-\gamma \dot{x}^2
\end{equation} for energy dissipation from such a fractional Lagrangian remained an open problem.

\section{The contour Caputo fractional derivative}\label{ii}
We here address the described problem of left and right derivatives. We would like to be able to freely swap the left and right derivatives. For this, we need to define a construction that is not sensitive to such changes.
\begin{figure}[H]
\centering
\begin{tikzpicture}

\coordinate (beg) at (-2,0);
\coordinate (t0up) at (0,1);
\coordinate (t1up) at (10,1);
\coordinate (t0down) at (0,-1);
\coordinate (t1down) at (10,-1);
\coordinate (end) at (12,0);

\draw[very thin,->,>=stealth] (beg) -- (end);

\draw[very thick,->] (t0down) -- (t1down);
\draw[very thick,<-] (t0up) -- (t1up);

\draw[thick,<-] (5,1) -- (6,1);      
\draw[thick,->] (5,-1) -- (6,-1);    

\draw (3.2,0.1) -- (3.2,-0.1);

\node at (-0.25, -0.25) {$t_0$};
\node at (10.25, -0.25) {$t_1$};
\node at (3.25, -0.28) {$t$};
\node at (12.25, -0.25) {$\mathbb{R}$};

\node at (5, 1.3) {upper interval};
\node at (5, -1.3) {lower interval};

\draw[dashed] (3.25,0.1) -- (10,0.1);
\draw[dashed] (3.25,-0.1) -- (10,-0.1);

\draw[<->] (2.25,0.05) -- (2.25,0.95);
\node[right] at (2.15,0.5) {$+i\varepsilon$};
\draw[<->] (2.25,-0.05) -- (2.25,-0.95);
\node[right] at (2.15,-0.5) {$-i\varepsilon$};
\end{tikzpicture}
\caption{Integration contour ${\Gamma_\varepsilon}\left([t_0,t_1]\right)$ with a chosen cut.}
\label{fig:contour}
\end{figure}
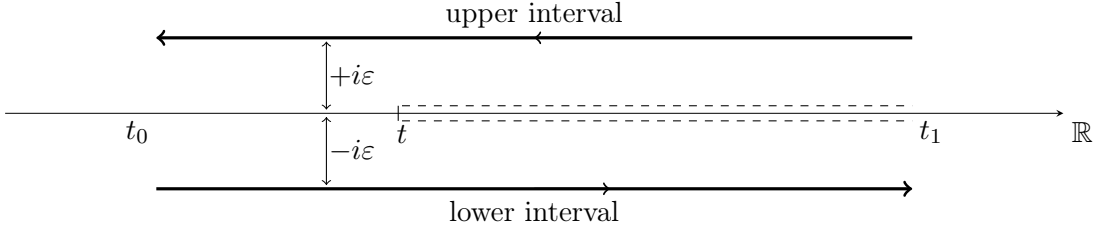

Let \(t_0,t_1\in\mathbb{R}\), \(t_0<t_1\), and let \(q\in C^2([t_0,t_1])\).
The dot always denotes differentiation with respect to the real variable:
\[
\dot q(t)=\frac{dq}{dt}(t), \qquad t\in[t_0,t_1].
\]
The complex variable appears only in the kernel of the contour integral.

\begin{definition}[Contour Caputo fractional derivative]
For \(t\in(t_0,t_1)\), define $q(\tau)\equiv q(\Re \tau)$,
\begin{equation}
{}_{}^{C}D_{\Gamma_\varepsilon}^{1/2}q(t)
\equiv
\frac{1}{\sqrt{\pi}}
\int\limits_{{\Gamma_\varepsilon}}
\frac{\dot q(\Re \tau)}{(\tau-t)^{1/2}}\,d\tau\bigg|_{\arg \mathbb{C} \in [-\pi,\pi)}.
\end{equation}
\end{definition}
Parametrizing the lower and upper branches (Fig.~\ref{fig:contour}) as
\[
\tau=s-i\varepsilon, \qquad \tau=s+i\varepsilon 
\]
and taking into account the orientation of the contour, we get
\[
{}^{C}D_{\Gamma_\varepsilon}^{1/2}q(t)
=
\frac{1}{\sqrt{\pi}}
\left[
\int_{t_0}^{t_1}
\frac{\dot q(s)}{(s-t-i\varepsilon)^{1/2}}\,ds
-
\int_{t_0}^{t_1}
\frac{\dot q(s)}{(s-t+i\varepsilon)^{1/2}}\,ds
\right].
\]

Now we split both integrals at the point \(s=t\):
\[
{}^{C}D_{\Gamma_\varepsilon}^{1/2}q(t)
=
\frac{1}{\sqrt{\pi}}
\left[
\int_{t_0}^{t}
\dot q(s)
\left(
\frac{1}{(s-t-i\varepsilon)^{1/2}}
-
\frac{1}{(s-t+i\varepsilon)^{1/2}}
\right)ds
\right.
\]
\[
\left.
+
\int_{t}^{t_1}
\dot q(s)
\left(
\frac{1}{(s-t-i\varepsilon)^{1/2}}
-
\frac{1}{(s-t+i\varepsilon)^{1/2}}
\right)ds
\right].
\]

Using the principal branch,
\[
\arg z\in[-\pi,\pi),
\]
we have, for \(s<t\),
\[
s-t-i0 \to |t-s|e^{-i\pi},
\qquad
s-t+i0 \to |t-s|e^{i\pi}.
\]
Therefore,
\[
\frac{1}{(s-t-i0)^{1/2}}\to\frac{i}{(t-s)^{1/2}},
\qquad
\frac{1}{(s-t+i0)^{1/2}}\to-\frac{i}{(t-s)^{1/2}}.
\]
Hence the contribution from \(s<t\) is
\[
\frac{2i}{\sqrt{\pi}}
\int_{t_0}^{t}
\frac{\dot q(s)}{(t-s)^{1/2}}\,ds.
\]

For \(s>t\), both boundary values coincide at $\lim\limits_{\varepsilon\to0}$:
\[
\frac{1}{(s-t-i0)^{1/2}}
=
\frac{1}{(s-t+i0)^{1/2}}
=
\frac{1}{(s-t)^{1/2}},
\]
so their difference vanishes. Thus,
\[
\lim_{\varepsilon\to 0^+}
{}^{C}D_{\Gamma_\varepsilon}^{1/2}q(t)
=
2i\,
\frac{1}{\sqrt{\pi}}
\int_{t_0}^{t}
\frac{\dot q(s)}{(t-s)^{1/2}}\,ds.
\]

In other words,
\[
\boxed{
\lim_{\varepsilon\to 0^+}
{}^{C}D_{\Gamma_\varepsilon}^{1/2}q(t)
=
2i\,{}^{C}_{t_0}D_t^{1/2}q(t),
}
\]
where
\[
{}^{C}_{t_0}D_t^{1/2}q(t)
=
\frac{1}{\sqrt{\pi}}
\int_{t_0}^{t}
\frac{\dot q(s)}{(t-s)^{1/2}}\,ds
\]
is the usual left-sided Caputo derivative of order \(1/2\).
\par Complex valued terms frequently appear in mathematical descriptions of
dissipative or open systems. For example, in an absorbing medium the wave
number may be written as
\begin{equation}
    k = k' + i k'',
\end{equation}
Similarly, one often introduces a complex dielectric permittivity,
\begin{equation}
    \varepsilon = \varepsilon' + i\varepsilon''.
\end{equation}
whose imaginary part describes electromagnetic absorption and energy
dissipation in the medium \cite{landau1984}.

Thus, the appearance of complex quantities is not unusual in the
descriptions of dissipative systems. Complex terms can encode
irreversible loss, while
the final physical observables remain real valued.
\section{Dissipation via Complex Root Branches}
\par The central mechanism relies on the branch sensitivity of the complexified time domain  $\sqrt{t}$. The action functional is defined with respect to an integration variable $dt$ constrained to a fixed branch of the argument,
\begin{equation}
    \arg \mathbb{C} \in [0,2\pi),
\end{equation}
thereby selecting a specific sheet of the associated Riemann surface: 
\begin{equation}
    S[q]
=
\int_{t_0}^{t_1}
L\!\left(
t,q(t),\dot q(t),\frac{d^\frac{1}{2}}{dt^\frac{1}{2}}q(t)
\right)\cdot d t \bigg|_{\arg \mathbb{C} \in [0,2\pi)} .
\end{equation}

In contrast, the fractional operators generating dissipation in medium involve an integration variable $d\tau$ which should exist on a distinct determination of the argument for the correct result,
\begin{equation}
\arg \mathbb{C} \in [-\pi,\pi),
\end{equation}
leading to a different branch of the same multi valued structure:
\begin{equation}
    \frac{d^\frac{1}{2}}{dt^\frac{1}{2}}q(t)\propto\int\frac{...}{\sqrt{\tau-t}}\cdot d\tau\bigg|_{\arg \mathbb{C} \in [-\pi,\pi)} .
\end{equation}

As a consequence, the conservative contribution to the action remains single branched in the chosen sheet, while dissipative terms arise from integrals whose support effectively probes a different branch of the complex contour.
\par We interpret this as follows: in the open system a point resides on a single branch of a complex root, and the principle of stationary action is defined for that point on its branch. At the same time, the Lagrangian contains a fractional term representing the interaction with the medium. However, the Lagrangian does not describe the medium as a whole; it describes only the behaviour of the point and the interaction between that point and the medium. Therefore, the medium’s complex time branch is different, but the interaction between the medium and the point is determined by the non-zero intersection of these branches with respect to the $4\pi$ periodicity:
$$
\text{point}\cap\text{medium},
$$
\begin{equation}
    [0,2\pi)\cap[-\pi,\pi)=[0,\pi).
\end{equation}
For the square root, the relevant periodicity of the argument is $4\pi$, since
\begin{equation}
    \exp\left(\frac{i(\theta+4\pi k)}{2}\right)
    =
    \exp\left(\frac{i\theta}{2}\right),
    \qquad k\in\mathbb{Z}.
\end{equation}
Therefore, the two intervals of arguments need not intersect as subsets of
$\mathbb{R}$ in order to define the same values of the square root.
It is sufficient that they intersect modulo $4\pi$.

For two intervals $A,B\subset\mathbb{R}$, we define their intersection
with respect to the $4\pi$ periodicity of the square root by
\begin{equation}
A\cap_{\sqrt{\phantom{z}}}B
:=
\left(A+4\pi\mathbb{Z}\right)
\cap
\left(B+4\pi\mathbb{Z}\right),
\end{equation}
where
\begin{equation}
A+4\pi\mathbb{Z}
:=
\bigcup_{k\in\mathbb{Z}}
\left(A+4\pi k\right).
\end{equation}

For example, although
\begin{equation}
[-5\pi,-3\pi)\cap[4\pi,6\pi)=\varnothing,
\end{equation}
the two intervals can be shifted by integer multiples of $4\pi$ without
changing the corresponding values of the square root. In particular,
\begin{equation}
{\theta+4\pi:\theta\in[-5\pi,-3\pi)}
=
[-\pi,\pi),
\end{equation}
while
\begin{equation}
{\theta-4\pi:\theta\in[4\pi,6\pi)}
=
[0,2\pi).
\end{equation}

Therefore,
\begin{equation}
[-5\pi,-3\pi)
\cap_{\sqrt{\phantom{z}}}
[4\pi,6\pi)
=
\left([-\pi,\pi)+4\pi\mathbb{Z}\right)
\cap
\left([0,2\pi)+4\pi\mathbb{Z}\right)
\
=
[0,\pi)+4\pi\mathbb{Z}.
\end{equation}

Choosing the representative in the interval $[-\pi,2\pi)$, this periodic
intersection reduces to
\begin{equation}
[-\pi,\pi)\cap[0,2\pi)
=
[0,\pi).
\end{equation}
The condition is therefore not
an ordinary intersection of the argument intervals, but their intersection
modulo the $4\pi$ periodicity of the square root.




\section{The principle of stationary action and the Euler-Lagrange equation}\label{v}

In this section we derive the Euler-Lagrange equation for a Lagrangian depending on the contour Caputo fractional derivative.

Let
\begin{equation}
L\equiv\frac{m\dot{q}^2(t)}{2}-U(q)+i\frac{\gamma}{8}\left({}_{}^{C}D_{\Gamma_\varepsilon}^{1/2}q(t)\right)^2,
\end{equation}

\begin{equation}\label{action_contour}
S[q]
=
\int_{t_0}^{t_1}
L\!\left(
t,q(t),\dot q(t),\frac{d^\frac{1}{2}}{dt^\frac{1}{2}}q(t)
\right)\,dt \bigg|_{\arg \mathbb{C} \in [0,2\pi)},
\end{equation}
where
\begin{equation}\label{def_w}
\frac{d^\frac{1}{2}}{dt^\frac{1}{2}}q(t)
:=
{}_{}^{C}D_{\Gamma_\varepsilon}^{1/2}q(t)
=
\frac{1}{\sqrt{\pi}}
\int\limits_{{\Gamma_\varepsilon}}
\frac{\dot q(\Re \tau)}{(\tau-t)^{1/2}}\,d\tau\bigg|_{\arg \mathbb{C} \in [-\pi,\pi)} .
\end{equation}

We consider variations
\[
q_\lambda(t)=q(t)+\lambda\eta(t),
\qquad \lambda\in\mathbb{R},
\]
where \(\eta\in C^1([t_0,t_1])\) satisfies
\begin{equation}\label{variation_conditions_rewritten}
\eta(t_0)=\eta(t_1)=0.
\end{equation}
Along the contour, we extend \(\eta\) by the rule
\[
\eta(\tau):=\eta(\Re\tau).
\]
Because \(\eta(t_0)=\eta(t_1)=0\), it follows that
\[
\eta(\tau)=0
\qquad
\text{on the vertical sides } \Re\tau=t_0 \text{ and } \Re\tau=t_1.
\]

The induced variation of \(w\) is
\begin{equation}
\delta \frac{d^\frac{1}{2}}{dt^\frac{1}{2}}q(t)
=
{}_{}^{C}D_{\Gamma_\varepsilon}^{1/2}\eta(t)
=
\frac{1}{\sqrt{\pi}}
\int\limits_{{\Gamma_\varepsilon}}
\frac{\dot\eta(\Re\tau)}{(\tau-t)^{1/2}}\,d\tau \bigg|_{\arg \mathbb{C} \in [-\pi,\pi)} .
\end{equation}

Therefore, the first variation of the action is
\begin{equation}\label{first_variation_full_rewritten}
\delta S
=
\int_{t_0}^{t_1}
\left[
\frac{\partial L}{\partial q}\,\eta
+
\frac{\partial L}{\partial \dot q}\,\dot\eta
+
\frac{\partial L}{\partial w}\,\delta w
\right]dt\bigg|_{\arg \mathbb{C} \in [0,2\pi)}.
\end{equation}

The classical term is treated by the usual integration by parts:
\begin{equation}
\int_{t_0}^{t_1}
\frac{\partial L}{\partial \dot q}\,\dot\eta\,dt
=
\left[
\eta\frac{\partial L}{\partial \dot q}
\right]_{t_0}^{t_1}
-
\int_{t_0}^{t_1}
\eta\,\frac{d}{dt}\!\left(\frac{\partial L}{\partial \dot q}\right)\,dt.
\end{equation}
Since \(\eta(t_0)=\eta(t_1)=0\), the boundary term vanishes, and 
\begin{equation}\label{classical_part_rewritten}
\int_{t_0}^{t_1}
\frac{\partial L}{\partial \dot q}\,\dot\eta\,dt
=
-
\int_{t_0}^{t_1}
\eta\,\frac{d}{dt}\!\left(\frac{\partial L}{\partial \dot q}\right)\,dt.
\end{equation}

\subsection{Fractional term}

We now consider the fractional contribution
\begin{equation}
I_f
:=
\int_{t_0}^{t_1}
\frac{\partial L}{\partial w}(t)
\left(
\frac{1}{\sqrt{\pi}}
\int\limits_{{\Gamma_\varepsilon}}
\frac{\dot\eta(\Re\tau)}{(\tau-t)^{1/2}}\,d\tau\bigg|_{\arg \mathbb{C} \in [-\pi,\pi)}
\right)dt\bigg|_{\arg \mathbb{C} \in [0,2\pi)}.
\end{equation}

For each fixed \(\varepsilon>0\), the kernel \((\tau-t)^{-1/2}\) is continuous on
\(\Gamma_\varepsilon\times[t_0,t_1]\), because the contour stays a positive distance away from the singular point \(\tau=t\).
Hence, the order of integration may be interchanged:
\begin{equation}
I_f
=
\frac{1}{\sqrt{\pi}}
\int\limits_{{\Gamma_\varepsilon}}
\dot\eta(\Re\tau)
\left(
\int_{t_0}^{t_1}
\frac{\partial L/\partial\frac{d^\frac{1}{2}}{du^\frac{1}{2}}q(u)}{(\tau-u)^{1/2}}\,du\bigg|_{\arg \mathbb{C} \in [0,2\pi)}
\right)d\tau\bigg|_{\arg \mathbb{C} \in [-\pi,\pi)}.
\end{equation}

Define
\begin{equation}
\Phi_{2\pi}(\tau)
\equiv
\int_{t_0}^{t_1}
\frac{\partial L/\partial\frac{d^\frac{1}{2}}{du^\frac{1}{2}}q(u)}{(\tau-u)^{1/2}}\,du\bigg|_{\arg \mathbb{C} \in [0,2\pi)}.
\end{equation}
Then,
\begin{equation}\label{If_phi_form_corrected}
I_f
=
\frac{1}{\sqrt{\pi}}
\int\limits_{{\Gamma_\varepsilon}}
\dot\eta(\Re\tau)\,\Phi_{2\pi}(\tau)\,d\tau\bigg|_{\arg \mathbb{C} \in [-\pi,\pi)}.
\end{equation}

\medskip

\noindent
\textbf{Lower interval.}
Here \(\tau=s-i\varepsilon\), with \(s:t_0\to t_1\), and \(d\tau=ds\). Hence,
\begin{equation}
I_f^{\mathrm{low}}
=
\frac{1}{\sqrt{\pi}}
\int_{t_0}^{t_1}
\dot\eta(s)\,\Phi_{2\pi}(s-i\varepsilon)\,ds\bigg|_{\arg \mathbb{C} \in [-\pi,\pi)}.
\end{equation}

\medskip

\noindent
\textbf{Upper interval.}
Here \(\tau=s+i\varepsilon\), with \(s:t_1\to t_0\), and \(d\tau=ds\). Hence,
\begin{equation}
I_f^{\mathrm{up}}
=
\frac{1}{\sqrt{\pi}}
\int_{t_1}^{t_0}
\dot\eta(s)\,\Phi_{2\pi}(s+i\varepsilon)\,ds\bigg|_{\arg \mathbb{C} \in [0,2\pi)}
=
-\frac{1}{\sqrt{\pi}}
\int_{t_0}^{t_1}
\dot\eta(s)\,\Phi_{2\pi}(s+i\varepsilon)\,ds\bigg|_{\arg \mathbb{C} \in [-\pi,\pi)}.
\end{equation}

\medskip

\noindent

We now integrate by parts in the real variable \(s\):
\begin{align}
\int_{t_0}^{t_1}
\dot\eta(s)\Bigl[\Phi_{2\pi}(s-i\varepsilon)-\Phi_{2\pi}(s+i\varepsilon)\Bigr]\,ds
&=
\left[
\eta(s)\Bigl(\Phi_{2\pi}(s-i\varepsilon)-\Phi_{2\pi}(s+i\varepsilon)\Bigr)
\right]_{t_0}^{t_1}
\nonumber\\
&\quad
-
\int_{t_0}^{t_1}
\eta(s)\,
\frac{d}{ds}
\Bigl[\Phi_{2\pi}(s-i\varepsilon)-\Phi_{2\pi}(s+i\varepsilon)\Bigr]\,ds\bigg|_{\arg \mathbb{C} \in [-\pi,\pi)}.
\end{align}
Since \(\eta(t_0)=\eta(t_1)=0\), the boundary term vanishes. Therefore,
\begin{equation}\label{If_after_parts_corrected}
I_f
=
-
\frac{1}{\sqrt{\pi}}
\int_{t_0}^{t_1}
\eta(s)\,
\frac{d}{ds}
\Bigl[\Phi_{2\pi}(s-i\varepsilon)-\Phi_{2\pi}(s+i\varepsilon)\Bigr]\,ds\bigg|_{\arg \mathbb{C} \in [-\pi,\pi)}.
\end{equation}

Thus, passing to the limit in order to collapse to the real axes $\mathbb{R}$, we obtain
\begin{equation}\label{If_final_jump_only}
\lim_{\varepsilon\to0^+} I_f
=
\frac{1}{\sqrt{\pi}}
\int_{t_0}^{t_1}
\eta(s)\,
\lim_{\varepsilon\to0^+}
\frac{d}{ds}\Bigl[
\Phi_{2\pi}(s+i\varepsilon)-\Phi_{2\pi}(s-i\varepsilon)
\Bigr]\,ds\bigg|_{\arg \mathbb{C} \in [-\pi,\pi)}.
\end{equation}

Combining this with \eqref{classical_part_rewritten}, the first variation becomes
\begin{equation}
\delta S
=
\int_{t_0}^{t_1}
\eta(t)
\left[
\frac{\partial L}{\partial q}
-
\frac{d}{dt}\left(\frac{\partial L}{\partial \dot q}\right)
+
\frac{1}{\sqrt{\pi}}
\lim_{\varepsilon\to0^+}
\frac{d}{dt}\Bigl[
\Phi_{2\pi}(t+i\varepsilon)-\Phi_{2\pi}(t-i\varepsilon)
\Bigr]
\right]dt\bigg|_{\arg \mathbb{C} \in [-\pi,\pi)}.
\end{equation}

Since \(\eta\) is arbitrary and satisfies \(\eta(t_0)=\eta(t_1)=0\), the fundamental lemma of the calculus of variations yields
\begin{equation}\label{EL_contour_jump_form}
\frac{1}{\sqrt{\pi}}\lim_{\varepsilon\to0^+}
\frac{d}{dt}\Bigl[
\Phi_{2\pi}(t+i\varepsilon)-\Phi_{2\pi}(t-i\varepsilon)
\Bigr]\bigg|_{\arg \mathbb{C} \in [-\pi,\pi)}
=\frac{d}{dt}\left(\frac{\partial L}{\partial \dot q}(t)\right)-\frac{\partial L}{\partial q}(t),
\qquad t\in(t_0,t_1)
.
\end{equation}

And \ref{EL_contour_jump_form} becomes (see Appendix~\ref{app:frictional_force}):
\begin{equation}
    -\gamma\dot{q}(t)=\frac{d}{dt}\left(\frac{\partial L}{\partial \dot q}(t)\right)-\frac{\partial L}{\partial q}(t).
\end{equation}

\section{Hamilton equations}\label{vi}
To obtain a physically meaningful Hamiltonian and corresponding equations of motion, it is essential to perform a Legendre transformation. In this context, we will treat the fractional derivative of $q$ as an independent projection, as it is not influenced by the velocity at the present time \( t \), but rather by the velocities at previous times. For more detailed arguments of such momentum definition see Appendix~\ref{app:Ostrogradsky-Riewe}:
\begin{equation}
    L(q,q^{\left(\frac{1}{2}\right)},\dot{q})=\frac{m\dot{q}^2}{2}+\frac{i\gamma \left( q^{\left(\frac{1}{2}\right)}\right)^2}{8}-U(q) .
\end{equation}
In order to make H independent of $\dot{q},q^{\left(\frac{1}{2}\right)}$ we will define canonical $p\equiv\frac{\partial L}{\partial \dot{q}},p_\frac{1}{2}\equiv\frac{\partial L}{\partial q^{\left(\frac{1}{2}\right)}}$:
\begin{equation}
    H(p,p_\frac{1}{2},q)\equiv p\dot{q}+p_\frac{1}{2} q^{\left(\frac{1}{2}\right)}-L(q,q^{\left(\frac{1}{2}\right)},\dot{q}),
\end{equation}
\begin{equation}
    dH=\dot{q}dp+q^{\left(\frac{1}{2}\right)}dp_\frac{1}{2}-\frac{\partial L}{\partial q}dq .
\end{equation}
Meanwhile,
\begin{equation}
    dH=\frac{\partial H}{\partial p}dp+\frac{\partial H}{\partial p_\frac{1}{2}}dp_\frac{1}{2}+\frac{\partial H}{\partial q}dq .
\end{equation}
Let us rewrite the Euler-Lagrange equation:
\begin{equation}
\frac{1}{\sqrt{\pi}}\lim_{\varepsilon\to0^+}
\frac{d}{dt}\Bigl[
\Phi_{2\pi}(t+i\varepsilon)-\Phi_{2\pi}(t-i\varepsilon)
\Bigr]\bigg|_{\arg \mathbb{C} \in [-\pi,\pi)}
=\frac{d}{dt}\left(\frac{\partial L}{\partial \dot q}(t)\right)-\frac{\partial L}{\partial q}(t),
\qquad t\in(t_0,t_1)
,
\end{equation}
where
\begin{equation}
\Phi_{2\pi}(\tau)
\equiv
\int_{t_0}^{t_1}
\frac{\partial L/\partial\frac{d^\frac{1}{2}}{du^\frac{1}{2}}q(u)}{(\tau-u)^{1/2}}\,du\bigg|_{\arg \mathbb{C} \in [0,2\pi)}\equiv\int_{t_0}^{t_1}
\frac{p_\frac{1}{2}(u)}{(\tau-u)^{1/2}}\,du\bigg|_{\arg \mathbb{C} \in [0,2\pi)}.
\end{equation}
\begin{definition}[Contour Riemann-Liouville fractional derivative]
For \(t\in(t_0,t_1)\), define $q(\tau)\equiv q(\Re \tau)$,
\begin{equation}
{}_{}D_{\Gamma_\varepsilon}^{1/2}p_\frac{1}{2}(t)
\equiv-
\frac{1}{\sqrt{\pi}}
\frac{d}{dt}\int\limits_{{\Gamma_\varepsilon}}
\frac{p_\frac{1}{2}(u)}{(t-u)^{1/2}}\,du\bigg|_{\arg \mathbb{C} \in [0,2\pi)}.
\end{equation}
\end{definition}
\begin{equation}
\lim_{\varepsilon\to0^+}
D_{\Gamma_\varepsilon}^{1/2}p_\frac{1}{2}(t)\bigg|_{\arg \mathbb{C} \in [-\pi,\pi)}
=\frac{d}{dt}\left(\frac{\partial L}{\partial \dot q}(t)\right)-\frac{\partial L}{\partial q}(t),
\qquad t\in(t_0,t_1)
.
\end{equation}
\begin{equation}
\lim_{\varepsilon\to0^+}
D_{\Gamma_\varepsilon}^{1/2}p_\frac{1}{2}(t)\bigg|_{\arg \mathbb{C} \in [-\pi,\pi)}
=\frac{d}{dt}p(t)+\frac{\partial H}{\partial q}(t),
\qquad t\in(t_0,t_1)
.
\end{equation}
Thus,
\begin{equation}
    \begin{cases}
        \frac{\partial H}{\partial p}=\dot{q} \\
        \frac{\partial H}{\partial p_\frac{1}{2}}=q^{\left(\frac{1}{2}\right)}\\
       \lim\limits_{\varepsilon\to0^+}
D_{\Gamma_\varepsilon}^{1/2}p_\frac{1}{2}(t)\bigg|_{\arg \mathbb{C} \in [-\pi,\pi)}
=\frac{d}{dt}p(t)+\frac{\partial H}{\partial q}(t),
\qquad t\in(t_0,t_1)
,
    \end{cases}
\end{equation}
\begin{equation}
    H(p,p_\frac{1}{2},q)=\frac{p^2}{2m}-\frac{2i}{\gamma}p_\frac{1}{2}^2+U(q).
\end{equation}

\section{Energy change}\label{vii}
With the inclusion of the new dissipative term in the Lagrangian, we observe that, in this case, the energy and the Hamiltonian do not coincide.
\begin{multline}
    \frac{dH}{dt}=\frac{\partial H}{\partial p} \frac{dp}{dt}+\frac{\partial H}{\partial p_\frac{1}{2}} \frac{dp_\frac{1}{2}}{dt}+\frac{\partial H}{\partial q} \frac{dq}{dt}= \\
    \frac{\partial H}{\partial p} \frac{dp}{dt}+\frac{\partial H}{\partial p_\frac{1}{2}} \frac{dp_\frac{1}{2}}{dt}+\left(-\frac{dp}{dt}+ \lim\limits_{\varepsilon\to0^+} 
D_{\Gamma_\varepsilon}^{1/2}p_\frac{1}{2}(t)\bigg|_{\arg \mathbb{C} \in [-\pi,\pi)}\right) \frac{dq}{dt}= \\ q^{\left(\frac{1}{2}\right)} \cdot\dot{p_\frac{1}{2}}+\dot{q}\cdot\lim\limits_{\varepsilon\to0^+}
D_{\Gamma_\varepsilon}^{1/2}p_\frac{1}{2}(t)\bigg|_{\arg \mathbb{C} \in [-\pi,\pi)} .
\end{multline}
Here we can define the anticommutator:
\begin{equation}\label{dH/dt1}
    \frac{dH}{dt}= \{q,p_\frac{1}{2}\}\equiv q^{\left(\frac{1}{2}\right)} \cdot\dot{p_\frac{1}{2}}+\dot{q}\cdot\lim\limits_{\varepsilon\to0^+}
D_{\Gamma_\varepsilon}^{1/2}p_\frac{1}{2}(t)\bigg|_{\arg \mathbb{C} \in [0,2\pi)}=-\frac{4i}{\gamma}p_\frac{1}{2}\dot{p_\frac{1}{2}}+\dot{q}(-\gamma\dot{q}) .
\end{equation}
From the other side we can introduce the energy $E$ as
\begin{equation}
    H(p,p_\frac{1}{2},q)=\frac{p^2}{2m}-\frac{2i}{\gamma}p_\frac{1}{2}^2+U(q)\equiv E-\frac{2i}{\gamma}p_\frac{1}{2}^2 ,
\end{equation}
\begin{equation}\label{dH/dt2}
     \frac{dH}{dt}=\frac{dE}{dt}-\frac{4i}{\gamma}p_\frac{1}{2}\dot{p_\frac{1}{2}} .
\end{equation}
By comparing \ref{dH/dt1} and \ref{dH/dt2} we can obtain the energy dissipation law:
\begin{equation}
    \frac{dE}{dt}=-\gamma\dot{q}^2 .
\end{equation}
\section{Conclusion}
By introducing a complex valued fractional term into the Lagrangian, we obtain a friction force proportional to the first derivative of the coordinate, derive the law of energy dissipation, and picture how energy is ``transferred'' along complex branches,  thereby providing an attempt for the geometric interpretation of this process in open systems.
\appendix

\section{The frictional force}\label{app:frictional_force}
\begin{equation}
L\equiv\frac{m\dot{q}^2(t)}{2}-U(q)+i\frac{\gamma}{8}\left({}_{}^{C}D_{\Gamma_\varepsilon}^{1/2}q(t)\right)^2
\end{equation}
Considering the fractional term after substituting the definition of \(\Phi_{2\pi}\) in \ref{EL_contour_jump_form}:
\begin{multline}    
    \frac{i\gamma}{8\pi}\cdot2\cdot\lim_{\varepsilon\to0^+}\Bigg[
\frac{d}{d\tau}
\int_{t_0}^{t_1}
\frac{1}{(\tau+i\varepsilon-u)^{1/2}}
\int\limits_{{\Gamma_\varepsilon}}
\frac{\dot q(\Re\sigma)}{(\sigma-u)^{1/2}}\,d\sigma\bigg|_{\arg \mathbb{C} \in [-\pi,\pi)}\,du\bigg|_{\arg \mathbb{C} \in [0,2\pi)}
\\
-
\frac{d}{d\tau}
\int_{t_0}^{t_1}
\frac{1}{(\tau-i\varepsilon-u)^{1/2}}
\int\limits_{{\Gamma_\varepsilon}}
\frac{\dot q(\Re\sigma)}{(\sigma-u)^{1/2}}\,d\sigma\bigg|_{\arg \mathbb{C} \in [-\pi,\pi)}\,du\bigg|_{\arg \mathbb{C} \in [0,2\pi)}
\Bigg]
\end{multline}

\[
=
\frac{i\gamma}{4\pi}\lim_{\varepsilon\to0^+}\Bigg[
\frac{d}{d\tau}
\int\limits_{{\Gamma_\varepsilon}}
\dot q(\Re\sigma)
\left(
\int_{t_0}^{t_1}
\frac{du}{(\tau+i\varepsilon-u)^{1/2}(\sigma-u)^{1/2}}\bigg|_{\arg \mathbb{C} \in [0,2\pi)}
\right)d\sigma\bigg|_{\arg \mathbb{C} \in [-\pi,\pi)}
\]
\begin{equation}
    -
\frac{d}{d\tau}
\int\limits_{{\Gamma_\varepsilon}}
\dot q(\Re\sigma)
\left(
\int_{t_0}^{t_1}
\frac{du}{(\tau-i\varepsilon-u)^{1/2}(\sigma-u)^{1/2}}
\bigg|_{\arg \mathbb{C} \in [0,2\pi)}\right)d\sigma\bigg|_{\arg \mathbb{C} \in [-\pi,\pi)}\Bigg]
\end{equation}
We can notice that:
\begin{equation}
    \frac{1}{(\tau\pm i\varepsilon-u)^{1/2}(\sigma-u)^{1/2}}
=
-2\,\frac{\partial}{\partial u}
\ln\!\Big((\tau\pm i\varepsilon-u)^{1/2}+(\sigma-u)^{1/2}\Big),
\end{equation}
Then:
\begin{equation}
\begin{aligned}
&=
\frac{i\gamma}{4\pi}\lim_{\varepsilon\to0^+}\Bigg[
\frac{d}{d\tau}
\int\limits_{{\Gamma_\varepsilon}}
\dot q(\Re\sigma)
\Bigg(
-2\ln\!\Big((\tau+i\varepsilon-t_1)^{1/2}\bigg|_{\arg \mathbb{C} \in [0,2\pi)}+(\sigma-t_1)^{1/2}\Big)
\\
&\hspace{5.2cm}
+\,2\ln\!\Big((\tau+i\varepsilon-t_0)^{1/2}\bigg|_{\arg \mathbb{C} \in [0,2\pi)}+(\sigma-t_0)^{1/2}\Big)
\Bigg)d\sigma\bigg|_{\arg \mathbb{C} \in [-\pi,\pi)}
\\
&\hspace{1cm}-
\frac{d}{d\tau}
\int\limits_{{\Gamma_\varepsilon}}
\dot q(\Re\sigma)
\Bigg(
-2\ln\!\Big((\tau-i\varepsilon-t_1)^{1/2}\bigg|_{\arg \mathbb{C} \in [0,2\pi)}+(\sigma-t_1)^{1/2}\Big)
\\
&\hspace{5.2cm}
+\,2\ln\!\Big((\tau-i\varepsilon-t_0)^{1/2}\bigg|_{\arg \mathbb{C} \in [0,2\pi)}+(\sigma-t_0)^{1/2}\Big)
\Bigg)d\sigma\bigg|_{\arg \mathbb{C} \in [-\pi,\pi)}
\Bigg].
\end{aligned}
\end{equation}
After $\frac{d}{d\tau}:$
\begin{equation}
\begin{aligned}
&=
\frac{i\gamma}{4\pi}\lim_{\varepsilon\to0^+}\Bigg[
\int\limits_{{\Gamma_\varepsilon}}
\dot q(\Re\sigma)
\Bigg(
-\frac{1}{
(\tau+i\varepsilon-t_1)^{1/2}
\Big((\tau+i\varepsilon-t_1)^{1/2}+(\sigma-t_1)^{1/2}\Big)}
\\
&\hspace{5.6cm}
+\frac{1}{
(\tau+i\varepsilon-t_0)^{1/2}
\Big((\tau+i\varepsilon-t_0)^{1/2}+(\sigma-t_0)^{1/2}\Big)}
\Bigg)d\sigma
\\
&\hspace{1cm}-
\int\limits_{{\Gamma_\varepsilon}}
\dot q(\Re\sigma)
\Bigg(
-\frac{1}{
(\tau-i\varepsilon-t_1)^{1/2}
\Big((\tau-i\varepsilon-t_1)^{1/2}+(\sigma-t_1)^{1/2}\Big)}
\\
&\hspace{5.6cm}
+\frac{1}{
(\tau-i\varepsilon-t_0)^{1/2}
\Big((\tau-i\varepsilon-t_0)^{1/2}+(\sigma-t_0)^{1/2}\Big)}
\Bigg)d\sigma
\Bigg].
\end{aligned}
\end{equation}

Multiplying numerator and denominator by the complex conjugate, we obtain:
\begin{equation}
\begin{aligned}
&=
\frac{i\gamma}{4\pi}\lim_{\varepsilon\to0^+}\Bigg[
\int\limits_{{\Gamma_\varepsilon}}
\dot q(\Re\sigma)
\Bigg(
-\frac{(\tau+i\varepsilon-t_1)^{1/2}-(\sigma-t_1)^{1/2}}{
(\tau+i\varepsilon-t_1)^{1/2}
(\tau+i\varepsilon-\sigma)}
+\frac{(\tau+i\varepsilon-t_0)^{1/2}-(\sigma-t_0)^{1/2}}{
(\tau+i\varepsilon-t_0)^{1/2}
(\tau+i\varepsilon-\sigma)
}\Bigg)d\sigma
\\
&\hspace{1cm}-
\int\limits_{{\Gamma_\varepsilon}}
\dot q(\Re\sigma)
\Bigg(
-\frac{(\tau-i\varepsilon-t_1)^{1/2}-(\sigma-t_1)^{1/2}}{
(\tau-i\varepsilon-t_1)^{1/2}
(\tau-i\varepsilon-\sigma)}
+\frac{(\tau-i\varepsilon-t_0)^{1/2}-(\sigma-t_0)^{1/2}}{
(\tau-i\varepsilon-t_0)^{1/2}
(\tau-i\varepsilon-\sigma)
}\Bigg)d\sigma
\Bigg].
\end{aligned}
\end{equation}

Simplifying:
\begin{equation}
\begin{aligned}
&
\frac{i\gamma}{4\pi}\lim_{\varepsilon\to0^+}\Bigg[
\int\limits_{{\Gamma_\varepsilon}}
\dot q(\Re\sigma)
\Bigg(
\frac{(\sigma-t_1)^{1/2}\bigg|_{\arg \mathbb{C} \in [-\pi,\pi)}}{
(\tau+i\varepsilon-t_1)^{1/2}\bigg|_{\arg \mathbb{C} \in [0,2\pi)}
(\tau+i\varepsilon-\sigma)}
-\frac{(\sigma-t_0)^{1/2}\bigg|_{\arg \mathbb{C} \in [-\pi,\pi)}}{
(\tau+i\varepsilon-t_0)^{1/2}\bigg|_{\arg \mathbb{C} \in [0,2\pi)}
(\tau+i\varepsilon-\sigma)
}\Bigg)d\sigma
\\
&\hspace{1cm}-
\int\limits_{{\Gamma_\varepsilon}}
\dot q(\Re\sigma)
\Bigg(
\frac{(\sigma-t_1)^{1/2}\bigg|_{\arg \mathbb{C} \in [-\pi,\pi)}}{
(\tau-i\varepsilon-t_1)^{1/2}\bigg|_{\arg \mathbb{C} \in [0,2\pi)}
(\tau-i\varepsilon-\sigma)}
-\frac{(\sigma-t_0)^{1/2}\bigg|_{\arg \mathbb{C} \in [-\pi,\pi)}}{
(\tau-i\varepsilon-t_0)^{1/2}\bigg|_{\arg \mathbb{C} \in [0,2\pi)}
(\tau-i\varepsilon-\sigma)
}\Bigg)d\sigma
\Bigg].
\end{aligned}
\end{equation}
\par For $t_0<\tau<t_1$ the multipliers below do not depend on $\sigma$ under the integral. Therefore, they can be taken outside the integral, and their limits can be evaluated separately.\\
For
\begin{equation}
    \arg\mathbb{C}\in[0,2\pi):
\end{equation}

\begin{equation}
    \begin{cases}
        \lim\limits_{\varepsilon\to0^+}(\tau+i\varepsilon-t_1)^\frac{1}{2}=\sqrt{t_1-\tau}\cdot\exp(i\frac{\pi}{2})=i\sqrt{t_1-\tau}, \\
       \lim\limits_{\varepsilon\to0^+}(\tau+i\varepsilon-t_0)^\frac{1}{2} =\sqrt{\tau-t_0}\cdot\exp(i\frac{0}{2})=\sqrt{\tau-t_0}\\
       \lim\limits_{\varepsilon\to0^+}(\tau-i\varepsilon-t_1)^\frac{1}{2}=\sqrt{t_1-\tau}\cdot\exp(i\frac{\pi}{2})=i\sqrt{t_1-\tau}\\
       \lim\limits_{\varepsilon\to0^+}(\tau-i\varepsilon-t_0)^\frac{1}{2} =\sqrt{\tau-t_0}\cdot\exp(i\frac{2\pi}{2})=-\sqrt{\tau-t_0}
    \end{cases}
\end{equation}
Then
\begin{equation}
\begin{aligned}
&=
\frac{i\gamma}{4\pi}\lim_{\varepsilon\to0^+}\Bigg[
\int\limits_{{\Gamma_\varepsilon}}
\dot q(\Re\sigma)
\frac{(\sigma-t_1)^{1/2}\bigg|_{\arg \mathbb{C} \in [-\pi,\pi)}}{
i\sqrt{t_1-\tau}\,(\tau+i\varepsilon-\sigma)}
\,d\sigma
\\
&\hspace{1cm}-
\int\limits_{{\Gamma_\varepsilon}}
\dot q(\Re\sigma)
\frac{(\sigma-t_0)^{1/2}\bigg|_{\arg \mathbb{C} \in [-\pi,\pi)}}{
\sqrt{\tau-t_0}\,(\tau+i\varepsilon-\sigma)}
\,d\sigma
\\
&\hspace{1cm}-
\int\limits_{{\Gamma_\varepsilon}}
\dot q(\Re\sigma)
\frac{(\sigma-t_1)^{1/2}\bigg|_{\arg \mathbb{C} \in [-\pi,\pi)}}{
i\sqrt{t_1-\tau}\,(\tau-i\varepsilon-\sigma)}
\,d\sigma
\\
&\hspace{1cm}-
\int\limits_{{\Gamma_\varepsilon}}
\dot q(\Re\sigma)
\frac{(\sigma-t_0)^{1/2}\bigg|_{\arg \mathbb{C} \in [-\pi,\pi)}}{
\sqrt{\tau-t_0}\,(\tau-i\varepsilon-\sigma)}
\,d\sigma
\Bigg].
\end{aligned}
\end{equation}

And

\begin{equation}
\begin{aligned}
&=
\frac{i\gamma}{4\pi}\lim_{\varepsilon\to0^+}\Bigg[
-\int\limits_{t_0}^{t_1}
\dot q(s)
\frac{(s+i\varepsilon-t_1)^{1/2}\bigg|_{\arg \mathbb{C} \in [-\pi,\pi)}}{
i\sqrt{t_1-\tau}\,(\tau-s)}
\,ds
+
\int\limits_{t_0}^{t_1}
\dot q(s)
\frac{(s-i\varepsilon-t_1)^{1/2}\bigg|_{\arg \mathbb{C} \in [-\pi,\pi)}}{
i\sqrt{t_1-\tau}\,(\tau-s+2i\varepsilon)}
\,ds
\\
&\hspace{1cm}
+\int\limits_{t_0}^{t_1}
\dot q(s)
\frac{(s+i\varepsilon-t_0)^{1/2}\bigg|_{\arg \mathbb{C} \in [-\pi,\pi)}}{
\sqrt{\tau-t_0}\,(\tau-s)}
\,ds
-
\int\limits_{t_0}^{t_1}
\dot q(s)
\frac{(s-i\varepsilon-t_0)^{1/2}\bigg|_{\arg \mathbb{C} \in [-\pi,\pi)}}{
\sqrt{\tau-t_0}\,(\tau-s+2i\varepsilon)}
\,ds
\\
&\hspace{1cm}
+\int\limits_{t_0}^{t_1}
\dot q(s)
\frac{(s+i\varepsilon-t_1)^{1/2}\bigg|_{\arg \mathbb{C} \in [-\pi,\pi)}}{
i\sqrt{t_1-\tau}\,(\tau-s-2i\varepsilon)}
\,ds
-
\int\limits_{t_0}^{t_1}
\dot q(s)
\frac{(s-i\varepsilon-t_1)^{1/2}\bigg|_{\arg \mathbb{C} \in [-\pi,\pi)}}{
i\sqrt{t_1-\tau}\,(\tau-s)}
\,ds
\\
&\hspace{1cm}
+\int\limits_{t_0}^{t_1}
\dot q(s)
\frac{(s+i\varepsilon-t_0)^{1/2}\bigg|_{\arg \mathbb{C} \in [-\pi,\pi)}}{
\sqrt{\tau-t_0}\,(\tau-s-2i\varepsilon)}
\,ds
-
\int\limits_{t_0}^{t_1}
\dot q(s)
\frac{(s-i\varepsilon-t_0)^{1/2}\bigg|_{\arg \mathbb{C} \in [-\pi,\pi)}}{
\sqrt{\tau-t_0}\,(\tau-s)}
\,ds
\Bigg].
\end{aligned}
\end{equation}
On this branch:
\begin{equation}
    \arg \mathbb{C}\in[-\pi,\pi), \qquad t_0<s<t_1
\end{equation}
\begin{equation}\label{pi_limits}
    \begin{cases}
        \lim\limits_{\varepsilon\to0^+}(s+i\varepsilon-t_1)^\frac{1}{2}=\sqrt{t_1-s}\cdot\exp(i\frac{\pi}{2})=i\sqrt{t_1-s}, \\
       \lim\limits_{\varepsilon\to0^+}(s+i\varepsilon-t_0)^\frac{1}{2} =\sqrt{s-t_0}\cdot\exp(i\frac{0}{2})=\sqrt{s-t_0}\\
       \lim\limits_{\varepsilon\to0^+}(s-i\varepsilon-t_1)^\frac{1}{2}=\sqrt{t_1-s}\cdot\exp(i\frac{(-\pi)}{2})=-i\sqrt{t_1-s}\\
       \lim\limits_{\varepsilon\to0^+}(s-i\varepsilon-t_0)^\frac{1}{2} =\sqrt{s-t_0}\cdot\exp(i\frac{0}{2})=\sqrt{s-t_0}
    \end{cases}
\end{equation}
After regrouping the terms:
\begin{equation}
\begin{aligned}
&=
\frac{i\gamma}{4\pi}\lim_{\varepsilon\to0^+}\Bigg[
-\int\limits_{t_0}^{t_1}
\dot q(s)
\frac{(s+i\varepsilon-t_1)^{1/2}\bigg|_{\arg \mathbb{C} \in [-\pi,\pi)}}{
i\sqrt{t_1-\tau}\,(\tau-s)}
\,ds
-
\int\limits_{t_0}^{t_1}
\dot q(s)
\frac{(s-i\varepsilon-t_1)^{1/2}\bigg|_{\arg \mathbb{C} \in [-\pi,\pi)}}{
i\sqrt{t_1-\tau}\,(\tau-s)}
\,ds
\\
&\hspace{1cm}
+\int\limits_{t_0}^{t_1}
\dot q(s)
\frac{(s+i\varepsilon-t_0)^{1/2}\bigg|_{\arg \mathbb{C} \in [-\pi,\pi)}}{
\sqrt{\tau-t_0}\,(\tau-s)}
\,ds
-
\int\limits_{t_0}^{t_1}
\dot q(s)
\frac{(s-i\varepsilon-t_0)^{1/2}\bigg|_{\arg \mathbb{C} \in [-\pi,\pi)}}{
\sqrt{\tau-t_0}\,(\tau-s)}
\,ds
\\
&\hspace{1cm}
+\int\limits_{t_0}^{t_1}
\dot q(s)
\frac{(s+i\varepsilon-t_1)^{1/2}\bigg|_{\arg \mathbb{C} \in [-\pi,\pi)}}{
i\sqrt{t_1-\tau}\,(\tau-s-2i\varepsilon)}
\,ds
+
\int\limits_{t_0}^{t_1}
\dot q(s)
\frac{(s-i\varepsilon-t_1)^{1/2}\bigg|_{\arg \mathbb{C} \in [-\pi,\pi)}}{
i\sqrt{t_1-\tau}\,(\tau-s+2i\varepsilon)}
\,ds
\\
&\hspace{1cm}
+\int\limits_{t_0}^{t_1}
\dot q(s)
\frac{(s+i\varepsilon-t_0)^{1/2}\bigg|_{\arg \mathbb{C} \in [-\pi,\pi)}}{
\sqrt{\tau-t_0}\,(\tau-s-2i\varepsilon)}
\,ds
-
\int\limits_{t_0}^{t_1}
\dot q(s)
\frac{(s-i\varepsilon-t_0)^{1/2}\bigg|_{\arg \mathbb{C} \in [-\pi,\pi)}}{
\sqrt{\tau-t_0}\,(\tau-s+2i\varepsilon)}
\,ds
\Bigg].
\end{aligned}
\end{equation}
With \ref{pi_limits} the first two rows go to zero.\\
We then proceed in a rigorous way by interpreting the integrals via the Sokhotski--Plemelj formula:
\begin{equation}
    \lim_{\varepsilon \to 0^+}
\left[
\int_{a}^{b} \frac{f(s)}{s - \tau - i\varepsilon}\,ds
-
\int_{a}^{b} \frac{f(s)}{s - \tau + i\varepsilon}\,ds
\right]
=
2\pi i\, f(\tau).
\end{equation}
\begin{equation}
\begin{aligned}
&=+0
\\
&\hspace{1cm}
+0
\\
&\hspace{1cm}
\frac{i\gamma}{4\pi}\lim_{\varepsilon\to0^+}\Bigg[
-\int\limits_{t_0}^{t_1}
\dot q(s)
\frac{(s+i\varepsilon-t_1)^{1/2}\bigg|_{\arg \mathbb{C} \in [-\pi,\pi)}}{
i\sqrt{t_1-\tau}\,(s-\tau+2i\varepsilon)}
\,ds
-
\int\limits_{t_0}^{t_1}
\dot q(s)
\frac{(s-i\varepsilon-t_1)^{1/2}\bigg|_{\arg \mathbb{C} \in [-\pi,\pi)}}{
i\sqrt{t_1-\tau}\,(s-\tau-2i\varepsilon)}
\,ds
\\
&\hspace{1cm}
-\int\limits_{t_0}^{t_1}
\dot q(s)
\frac{(s+i\varepsilon-t_0)^{1/2}\bigg|_{\arg \mathbb{C} \in [-\pi,\pi)}}{
\sqrt{\tau-t_0}\,(s-\tau+2i\varepsilon)}
\,ds
+
\int\limits_{t_0}^{t_1}
\dot q(s)
\frac{(s-i\varepsilon-t_0)^{1/2}\bigg|_{\arg \mathbb{C} \in [-\pi,\pi)}}{
\sqrt{\tau-t_0}\,(s-\tau-2i\varepsilon)}
\,ds
\Bigg].
\end{aligned}
\end{equation}
\begin{equation}
    =-\gamma\dot{q}(\tau)
\end{equation}
\section{Remark on the Ostrogradsky-Riewe constuction of the higher-order momentum}\label{app:Ostrogradsky-Riewe}
We attempted to apply Ostrogradsky's construction 
to our fractional Lagrangian as if it were a higher-order system. We also note that Fred Riewe tried a similar (''Ostrogradsky's") approach \cite{riewe1996}, but in doing so introduced an unphysical time limit and changed the order of integration without sufficient justification.
Recall that in Ostrogradsky’s approach \cite{ostrogradsky1850,harmanni2016}, 
the equation of motion is written as:
\[
\sum_{i=0}^n (-1)^i \,\frac{d^i}{dt^i} 
\Bigl(\,\frac{\partial L}{\partial q^{(i)}}\Bigr) 
\;=\; 0,
\]
which contains terms up to \(q^{(2n)}\). For the \(2n\) canonical 
coordinates, Ostrogradsky’s choices are:
\[
Q_i \;\equiv\; q^{(i-1)},
\quad
P_i \;\equiv\; \sum_{j=i}^{n}\,\Bigl(-\frac{d}{dt}\Bigr)^{j-i}\,
\frac{\partial L}{\partial q^{(j)}}.
\]
These definitions lead to the Hamiltonian
\[
H \;\equiv\; \sum_{i=1}^n\,P_i\,q^{(i)} \;-\; L.
\]

In essence, the definition of momentum in ``higher-order''/fractional contexts 
amounts to ``factoring out'' the differentiation operator from the equations of motion.
However, in our fractional setting, there are multiple ways to factorize or partition 
the fractional derivatives, and none of those 
variants yielded consistent Hamilton equations for this model. The core difficulty is 
that in a fractional framework, the straightforward ``Ostrogradsky factorization'' 
fails to yield a meaningful representation of the system’s dynamics. By contrast, the simpler 
procedure adopted here—treating the fractional derivative \(q^{\left(\frac{1}{2}\right)}\) as an independent 
projection—does produce a coherent set of equations of motion and a physically interpretable 
Hamiltonian.

\end{document}